\documentstyle[12pt]{article} 
\normalbaselines
\begin{document}
\bibliographystyle{unsrt}


\noindent LA-UR-97-2760 

\begin{center}
{\large \bf THE DISCOVERY OF SQUEEZED STATES \\[2mm]
 --- IN 1927}\\[7mm]
Michael Martin Nieto\footnote{Email: mmn@pion.lanl.gov} \\
{\it Theoretical Division (T-8, MS-B285), 
Los Alamos National Laboratory, \\
University of California,
Los Alamos, New Mexico 87545 U.S.A.}\\
and\\
{\it Universit\"at Ulm, 
Abteilung f\"ur Quantenphysik, \\
Albert-Einstein-Allee 11, 
D 89069 Ulm, 
GERMANY}\\[5mm]
\end{center}

\vspace{2mm}

\begin{abstract}
I first review a) the flowering of coherent states in the 1960's, yet
b) the discovery of coherent states in 1926, and c) the flowering of 
squeezed states in the 1970's and 1980's.  Then, with the background of the
excitement over the then new quantum mechanics, I describe d) the discovery 
of squeezed states in 1927.
\end{abstract}


\section{Coherent States}

Coherent states are important in many fields of 
physics \cite{1,2}.  This became widely recognized during
the 1960's due to the
work of Glauber \cite{glauber}, Klauder \cite{klauder,ks}, and 
Sudarshan \cite{ks,sudar}.

In modern parlance, they are standardly defined in three equivalent ways:

1) {\it Displacement-Operator Method}. For the harmonic oscillator, coherent
states, $|\alpha \rangle$, are given by the unitary displacement 
operator acting on the ground state:
\begin{equation}
        D(\alpha)|0\rangle  =           
\exp\left[-\frac{1}{2}|\alpha|^2\right] \sum_{n} \frac{\alpha ^n}{\sqrt{n!}} 
|n\rangle  
\equiv |\alpha\rangle  ,  \label{D} 
\end{equation}
\begin{equation}
D(\alpha)= \exp[\alpha a^{\dagger} - \alpha^*a]
     = \exp[-|\alpha|^2/2]\exp[\alpha a^{\dagger}] \exp[-\alpha^* a]~.
\end{equation} 
The generalization of this method to arbitrary Lie groups has an 
involved history.\footnote{Unknown to many, Klauder 
developed this method in an early paper \cite{kdop}.} 
(See Refs. \cite{kdop,dop} and later reviews \cite{1,2}.)   
One simply applies the 
generalized displacement operator, which is the  unitary
exponentiation of the factor algebra, on to an extremal state.

2) {\it Ladder- (Annihilation-) Operator Method}.  For the harmonic oscillator, 
the coherent states are also the eigenstates of the destruction operator:
\begin{equation}
a|\alpha\rangle = \alpha |\alpha\rangle.
\end{equation}
That these states are the same as the displacement-operator states
follows from Eq. (\ref{D}), since
\begin{equation} 
0 = D(\alpha)a|0\rangle =  (a - \alpha)D(\alpha)|0\rangle =
(a - \alpha)|\alpha \rangle .
\end{equation}
The generalization to arbitrary Lie groups is straight forward, and 
has also been studied \cite{1,2}. 

3) {\it Minimum-Uncertainty Method}.  This method, as we will see in 
the next section, intuitively harks back to 
Schr\"{o}dinger's discovery of the coherent states \cite{schcs}.   

\indent For the harmonic oscillator, 
the classical variables $x$ and $p$
vary as the $\sin$ and the $\cos$ of the classical $\omega t$.  
If one then takes these ``natural" classical
variables and transforms them into quantum operators, they
define a commutation relation and an uncertainty relation: 
\begin{equation} 
[x,p] = i, \hspace{0.5in} 
(\Delta x)^2(\Delta p)^2 \geq {\frac{1}{4}}.  \label{uncert}
\end{equation}
The states that minimize this uncertainty relation are given by the 
solutions to the equation
\begin{equation}
\left(x + \frac{i }{2(\Delta p)^2} p\right)\psi_{mus}
=\left(\langle x\rangle +\frac{i }{2(\Delta p)^2}\langle p\rangle
\right)\psi_{mus}.  \label{cond}
\end{equation}
[Of the  four parameters $\langle x\rangle$, 
$\langle x^2\rangle$, $\langle p\rangle$, and 
$\langle p^2\rangle $, only three are
independent because  the equality in the uncertainty relation 
is satisfied.]  Rewrite Eq. (\ref{cond}) as 
\begin{equation}
\left(x + iB p\right)\psi_{mus} = C \psi_{mus}  ,~~~~~~~~~~
B = \frac{\Delta x}{\Delta p},  ~~~~~~~~~~
 C = \langle x\rangle + i B \langle p\rangle .
\end{equation}
Here $B$ is real and $C$ is complex.  These states, $\psi_{mus}(B,C)$, are the 
minimum-uncertainty states.   

$B$ can be adjusted to $B_0$ 
so that the ground-eigenstate is a member of the set.  
Then these restricted states, $\psi_{mus}(B=B_0,C)=\psi_{cs}(B_0,C)$,
are the minimum-uncertainty coherent states:
\begin{equation}
\psi_{cs} = \pi^{-1/4}\exp\left[-\frac{(x-x_0)^2}{2}+ip_0x\right]~.
\label{sch}
\end{equation}
By the use of the Hermite polynomial generating function, and
with the identifications
\begin{equation}
\alpha=\alpha_1+i\alpha_2= \frac{x_0+ip_0}{\sqrt{2}}~,
\end{equation}
it can be shown that Eqs.  (\ref{D}) and (\ref{sch}) are equivalent. 
This method has been applied to
general Hamiltonian potential 
and symmetry systems, yielding generalized coherent states
\cite{n1}.  

The coherent wave packets are Gaussians, with widths that of the ground-state
Gaussian.  With time, the centroid of the packet follows the classical 
motion and does not change its shape.


\section{The Discovery of Coherent States}

The first half of 1926 was an amazingly productive period for 
Erwin Schr\"odinger.  He submitted six important papers \cite{schcs},
beginning with his fundamental paper solving the hydrogen-atom.    

There was, of course, tremendous controversy surrounding Schr\"odinger's 
work.  In particular Lorentz wrote to Schr\"odinger on May 27 lamenting 
the fact that his wave functions were stationary, and 
did not display classical motion.  On June 6 Schr\"odinger replied that 
he had found a system where classical motion was seen, and sent Lorentz
a draft copy of the paper we are interested in, 
{\it Der stetige \"Ubergang von der Mikro- zur Makromechanik} \cite{schcs}.

Using  the  generating function found
in the classic book of Courant and Hilbert, Schr\"odinger realized  
that a Gaussian wave-function could be constructed from a particular 
superposition of the wave functions corresponding to the discrete 
eigenvalues of the harmonic oscillator.  Further, these new states 
followed the classical motion.  At this time the 
probability-{\it amplitude} nature of the wave function was not yet 
known, so the complex nature of the wave function bothered Schr\"odinger.
He wondered if, perhaps, it was only the real part of the wave 
function that is physical.  

But in any event, even though the uncertainty relation had yet to be 
discovered, from a view point that most closely resembles 
the modern minimum-uncertainty method, Schr\"odinger had discovered
the coherent states. 

During this period there was great rivalry between Schr\"odinger and 
the ``young Turks," Heisenberg and Pauli.  Indeed, Pauli felt it necessary
to mollify Schr\"odinger after having called his views, 
``local Z\"urich superstitions."\footnote{{\it Z\"uricher 
Lokalaberglauben}} Heisenberg was even stronger, saying
Schr\"odinger's physical picture makes scarcely any sense, ``..in other
words, I think it is crap."\footnote{{\it in a. W. ich finde es Mist.}}
 --- Then again, Werner was hard to please.  He later called the 
Dirac theory, ``learned crap."\footnote{{\it gelehrten Mist}} 
(See Ref. \cite{paul}.)

This period amusingly came to an end when Born corrected himself in 
proof.  He realized that the probability was 
not proportional to the wave function, but to its square.


\section{Squeezed States}

Squeezed states were mathematically rediscovered and discussed by
many authors \cite{ss1}, 
especially in the 1970's and early 80's.  Each author
tended to emphasize differing aspects of the system.  Interest grew in 
squeezed states as it began to appear that experimentalists might be 
able to observe them \cite{nato}. 
This was first true in the field of 
gravitational-wave detection \cite{caves}.    
(Indeed, the term ``squeezed states" was
coined in the context of Weber-bar gravitational-wave detection \cite{hol}.)
But, especially with the quantum-optics experimental breakthroughs 
starting in the mid-1980's \cite{issues}, 
the field exploded, as evidenced also by 
the present conference series.

As coherent states, the more general squeezed states can be defined
in three equivalent ways:

1) {\it Displacement-Operator Method}.
To obtain squeezed states, one applies both the squeeze and displacement
operators onto the ground state: 

\begin{equation}
D(\alpha)S(z)|0\rangle = |\alpha,z\rangle~, 
   ~~~~~~  z = re^{i\phi} = z_1 + i z_2~,    ~~~~~~
S(z) = \exp\left[\frac{1}{2}za^{\dagger}a^{\dagger}-\frac{1}{2}z^*aa\right]~, 
\label{xx}
\end{equation}
\begin{equation}
S(z)  =   \exp\left[{\frac{1}{2}}e^{i\phi}(\tanh r)
a^{\dagger}a^{\dagger}\right]
\left({\frac{1}{\cosh r}}\right)^{({{\frac{1}{2}}+a^{\dagger}a})}
\exp\left[-{\frac{1}{2}}e^{-i\phi}(\tanh r)aa\right].  \label{b} 
\end{equation}
The ordering of $DS$ vs. $SD$ in Eq. (\ref{xx}) is equivalent, 
amounting to a change of parameters.  

2) {\it Ladder- (Annihilation-) Operator Method}.  Using a 
Holstein-Primakoff/Bogoliubov transformation, $DSaS^{-1}D^{-1}$ 
is a linear combination of $a$ and $a^{\dagger}$, that yields
\begin{equation}
\left[(\cosh r) a -(e^{i\phi}\sinh r) a^{\dagger}\right]
|\alpha,z\rangle = 
\left[(\cosh r) \alpha -(e^{i\phi}\sinh r) \alpha^*\right]. 
|\alpha,z\rangle
\end{equation}

3) {\it Minimum-Uncertainty Method}.  If one takes the convention 
$z=r ={\it Real}$, which is valid because $\phi$ amounts to
an initial time, $t_0$, one can return to Eq. (\ref{cond}) and realize that
the $\psi_{mus}$ are squeezed states.  To include the phase explicitly, 
one moves a step further \cite{nagel} 
than the Heisenberg uncertainty relation to the 
Schr\"odinger uncertainty relation \cite{schuncert}:
\begin{equation}
(\Delta x)^2(\Delta p)^2 \geq
|\langle [x,p]/2 + \{x,p\}/2\rangle|^2.  \label{suncert}
\end{equation}
The states which satisfy this uncertainty relation, and which can be shown to 
be equivalent to the other formulations \cite{sns}, are 
\begin{equation}
\psi_{ss}    
=  \frac{e^{-ix_0p_0/2}}{\pi^{1/4}[{\cal F}_1]^{1/2}} 
     \exp\left[-\frac{(x-x_0)^2}{2}{\cal F}_2+ip_0x\right] ,  \label{sn}
\end{equation}
\begin{eqnarray}
{\cal F}_1 &=& 
         \cosh r +e^{i\phi}\sinh r,   ~~~~~~~
{\cal F}_2 
    = \frac{\cosh r - e^{i\phi} \sinh r} 
           {\cosh r + e^{i\phi} \sinh r},      
\end{eqnarray}
With time, the centroids of these wave packets also follow the 
classical motion, but they do not retain their shapes.  The 
width of a particular Gaussian oscillates as 
\begin{equation}
[\Delta x(t)]^2 = 
[(\cosh r)^2 + (\sinh r)^2 +2(\cosh r)(\sinh r)\cos(2t -\phi)] , 
         \label{dx}
\end{equation}
\begin{equation}
[\Delta p(t)]^2 = 
[(\cosh r)^2 + (\sinh r)^2 -2(\cosh r)(\sinh r)\cos(2t -\phi)] ,  
\end{equation}
\begin{equation}
4{[\Delta x(t)]^2[\Delta p(t)]^2}
=  1 +\frac{1}{4}\left(s^2-\frac{1}{s^2}\right)^2\sin^2(2t-\phi) ,~
~~~~~~s = e^r.    \label{dxdp}
\end{equation}

Generalizations have been given for other systems \cite{gss,trif}.
Further, from Eq. (\ref{suncert}), 
one can realize that generalized intelligent states \cite{trif} 
are, in fact, generalized squeezed states.


\section{The Discovery of Squeezed States}

In 1926 a Cornell University assistant professor, Earle Hesse Kennard 
(born 1885), was granted a sabbatical.  Upon his return he would be
would be promoted to full professor.  In October Kennard arrived at
Max Born's Institut f\"ur Theoretische Physik of the University of 
G\"ottingen, where Heisenberg and  Jordan  had also worked.
There Kennard mastered the matrix mechanics of Heisenberg 
and the wave mechanics of Schr\"odinger.

At the same time, Heisenberg submitted his uncertainty relation paper 
and went to Copenhagen to work with Bohr.  By the spring of 1927, Bohr 
was working on his famous Como talk on the uncertainty relation, which 
later became a special supplement to Nature.  
Kennard moved to Copenhagen on March 7.  While there 
he completed the manuscript in which squeezed states were discovered
\cite{kennard}, {\it Zur Quantenmechanik einfacher Bewegungstypen}.  
Kennard acknowledges the help of Bohr and Heisenberg in the paper.  
Also, in the grand European-professor tradition, Kennard left the 
paper with  Bohr for Bohr to review and approve it,  and 
then to submit it \cite{bohrarc}.

The paper itself begins with a review of matrix and wave mechanics. 
Then Kennard 
discusses the time-evolution of three systems: i) a particle in an 
electric field, ii) a particle in a magnetic field, and, so important to us, 
in Sec. 4C iii) a general Gaussian in a harmonic oscillator potential.

Kennard observes  that the classical motion is followed and, 
in his own notation, gives the critical Eqs. (\ref{dx})-(\ref{dxdp}) above.  
He also explains their meaning, i.e., that a) the sum of the uncertainties in 
x and p is a constant, b) the uncertainty product varies from the 
minimum as $\sin^22t$ [$\sin^2t\cos^2t$], c) the product has a minimum only 
twice each half period,\footnote{Kennard wrote, {\it nur zweimal in jeder 
Schwingung}, so by  {\it Schwingung} he meant ``swing" instead of 
an entire oscillation.}  and d) when the uncertainties are equal 
(in natural units) then the (coherent) states of Schr\"odinger 
\cite{schcs} are obtained.  In other words, he had everything.

One can speculate why the papers of Schr\"odinger \cite{schcs} and 
Kennard \cite{kennard} were, in the main, ignored for so long.  I think 
the simple answer is that they were too far ahead of their time.  During 
the 20's, most would have felt 
it was inconceivable that this work meant more than a matter of 
principle.  It would be decades before connection was made to experiment. 

To be popular 
in physics you have to either be good or lucky.  
Sometimes it is better to be lucky.  
But if you are going to be good, 
perhaps you shouldn't  be too good. 

A more detailed 
version of this discussion will appear elsewhere \cite{kennardme}.


\section*{Acknowledgments}

Among all the people who have helped me on this, I would especially like
to thank Jarman Kennard and Felicity Pors.
This work was supported by the U.S. Department of 
Energy and the Alexander von Humboldt Foundation.  



\begin{thebibliography}{99}

\bibitem{1} 
J.~R.~Klauder and B.-S. Skagerstam, {\it Coherent States -- 
Applications in Physics
and Mathematical Physics} (World Scientific, Singapore, 1985);.

\bibitem{2}
W.-M. Zhang, D. H. Feng,
and R. Gilmore, Rev. Mod. Phys. {\bf 62}, 867 (1990).

\bibitem{glauber}  R. J. Glauber, Phys. Rev. {\bf 130}, 2529 (1963).

\bibitem{klauder}  J. R. Klauder, Annals of Physics
{\bf11}, 123 (1960); J. Math. Phys. {\bf 4}, 1055 (1963);
J. Math. Phys. {\bf 4}, 1058 (1963).  

\bibitem{ks} 
J. R. Klauder and E. C. G. Sudarshan, {\it Fundamentals of
Quantum Optics} (Benjamin, NY, 1968). 

\bibitem{sudar}
E. C. G. Sudarshan, 
Phys. Rev. Lett. {\bf 10}, 227 (1963). 

\bibitem{kdop}
J. R. Klauder, J. Math. Phys. {\bf 4}, 1058 (1963).   

\bibitem{dop}
A. O. Barut and L. Girardello, Commun.\
Math.\ Phys.\ {\bf 21}, 41 (1971); 
A. M. Perelomov, Commun.\ Math.\ Phys.\ {\bf {26}}, 222 (1972); 
{\it Generalized Coherent States and Their Applications} 
(Springer-Verlag, Berlin, 1986); 
R. Gilmore, {\it Lie Groups, Lie Algebras, 
and Some of Their Applications} (Wiley, New York, 1974). 

\bibitem{schcs}
E. Schr\"odinger, Naturwiss. {\bf 14}, 664 (1926).

\bibitem{n1}    M. M. Nieto and L. M.
Simmons, Jr., Phys. Rev. Lett. {\bf 41}, 207 (1978); Phys.\ Rev.\ D {\bf 20}, 
1321 (1979), the first of a series concluding with M. M. Nieto, L. M. 
Simmons, Jr., and V. P. Gutschick, Phys.\ Rev.\ D {\bf 23},  927 (1981).  
M. M. Nieto, in  Ref. \cite{1}, p.\ 429, gives a summary of this program. 

\bibitem{sch} E. Schr\"odinger, {\it Collected Papers on Wave Mechanics}
(Blackie \& Son, London, 1928).  

\bibitem{paul} W. Pauli, {\it Scientific Correspondence}
(Springer, New York, Vol I-1979, Vol II-1986);  letter 147 dated 22 Nov
1926, letter 136 dated 8 June 1926, letter 353 dated 8 Feb 1926.

\bibitem{ss1}  H. Takahasi, Ad. Communication Systems {\bf 1},
227 (1965);
D. Stoler, Phys. Rev. D {\bf 1}, 3217 (1970);
ibid. {\bf 4}, 1925 (1974);
H. P. Yuen, Phys. Rev. A {\bf 13}, 2226 (1976);
J. N.  Hollenhorst,  Phys. Rev. D {\bf 19}, 1669 (1979);
I. Fujiwara and K. Miyoshi, Prog. Theor. Phys. {\bf 64}, 
715 (1980);
V. V. Dodonov, E. V. Kurmyshev, and V. I. Man'ko, 
Phys. Lett. A {\bf 79}, 150 (1980);
P. A. K. Rajogopal and J. T. Marshall, Phys. Rev. A 
{\bf 26}, 2977 (1982);
H. P. Yuen, Phys. Rev. Lett. {\bf 51}, 719 (1983).

\bibitem{hol} J. N. Hollenhorst, in Ref. \cite{ss1}.

\bibitem{nato}  For a pedagogical review from this period, see
M. M. Nieto, in: {\em Frontiers of Nonequilibrium 
Statistical Physics}, eds.  G. T. Moore and M. O. Scully (Plenum, New
York, 1986) p. 287. 

\bibitem{caves} C. M. Caves, J. S. Thorne, R. W. P. Drever, V. D. 
Sandberg, and M. Zimmerman, Rev. Mod. Phys. {\bf 52}, 341 (1980).

\bibitem{issues}  See the special issues: 
J. Opt. Soc. Am. B {\bf 4}, No. 10 (1987); 
J. Mod. Opt. {\bf 34}, No. 6 (1987); 
Appl. Phys. B {\bf 55}, No 3 (1992).

\bibitem{nagel}  B. Nagel, in: {\it Modern Group Theoretical Methods 
in Physics}, ed. J. Bertrand, et al. (Kluwer, NL, 1995), p. 221.  

\bibitem{schuncert} E. Schr\"odinger, Sitzungb. Preuss. Akad. Wiss., 
221 (1930).  

\bibitem{sns} M. M. Nieto, Phys. Lett. A {\bf 229}, 135 (1997).

\bibitem{gss} M. M. Nieto and D. R. Truax, Phys. Rev. Lett. 
{\bf 71}, 2843 (1993).

\bibitem{trif} D. A. Trifonov, J. Math. Phys. {\bf 35}, 2297 (1994).

\bibitem{kennard} E. H. Kennard, Zeit. Phys. {\bf 44}, 326 (1927).  
See Sec. 4C.

\bibitem{bohrarc}  Niels Bohr Archives, Bohr Scientific Correspondence, 
microfilm 13, five letters between Bohr and Kennard.

\bibitem{kennardme} M. M. Nieto, in preparation.

\end {thebibliography}
\end{document}